\documentclass[showpacs,twocolumn,prl]{revtex4}
\usepackage{amssymb}
\usepackage{graphicx}

\begin{document}

\title{Spin Hall Effect in a Doped Mott Insulator}
\author{Su-Peng Kou$^{1,2},$ Xiao-Liang Qi$^{3}$ and Zheng-Yu Weng$^{3}$}
\affiliation{$^{1}$Department of Physics, Massachusetts Institute of Technology,
Cambridge MA 02139, USA\\
$^{2}$Department of Physics, Beijing Normal University, Beijing 100875, China%
\\
$^{3}$Center for Advanced Study, Tsinghua University, Beijing 100084, China}

\begin{abstract}
The existence of conserved spin Hall currents is shown in a strongly
correlated system without involving spin-orbit coupling. The spin Hall
conductivity is determined by intrinsic bulk properties, which remains
finite even when the charge resistivity diverges in strong magnetic fields
at zero temperature. The state in the latter limit corresponds to a spin
Hall insulator. Such a system is a doped Mott insulator described by the
phase string theory, and the spin Hall effect is predicted to coexist with
the Nernst effect to characterize the intrinsic properties of the
low-temperature pseudogap phase.
\end{abstract}

\maketitle

\section{Introduction}

Recently, the proposals \cite{Mur1,niu} of manipulating spin currents by an
electric field via spin Hall effect (SHE) have attracted a lot of attention,
in which studies have been mainly focused on non-interacting or weakly
interacting semiconductor systems with substantial spin-orbit (SO) coupling.
Due to the SO coupling, an applied electric field can induce intrinsic spin
Hall currents in a transverse direction, which can be well separated from
the dissipative charge currents along the longitudinal direction.

But the intrinsic SHE in an SO system is usually very sensitive to disorder
\cite{diss,sheng} because the spin polarization is tied to momentum and the
spin current is not conserved. Without the conservation law, the spin Hall
conductivity studied in the linear response theory generally may not be
directly connected to physical spin accumulations in such SO systems \cite%
{sheng,zhang}, and a disorder effect can drastically affect the latter,
leading to the debate whether an intrinsic SHE can meaningfully exist in
realistic systems.

It is thus a very interesting issue whether an intrinsic SHE can exist in
electron systems without the SO coupling such that the conservation law
still holds for the spin current and the spin transport becomes
well-described. Without the SO coupling, however, the difficulty is how to
effectively couple an electric field to a neutral spin current detached from
the charge current. In other words, one needs to identify a system in which
the spin current is not directly carried by the dissipative charge current,
and at the same time the former can be directly driven by the electric
field. For a weakly interacting system where elementary excitations are
quasiparticles which carry both charge and spin under a Fermi-Liquid
description, the spin and charge currents usually cannot be simply separated
to find such kind of ``dissipationless'' spin Hall effect.

On the other hand, in the \emph{strongly correlated} electron systems
related to the high-$T_{c}$ cuprates, a (partial) separation of spin and
charge degrees of freedom \cite{anderson} makes it possible to manipulate
conserved spin currents which may be distinguishable from ordinary
dissipative charge currents. In this paper, we shall make a proposal that
``non-dissipative''\ spin Hall currents indeed exist in the so-called lower
pseudogap phase (LPP) of the cuprates under the description of the phase
string theory \cite{sv}. To be sure, so far there is no consensus concerning
the correct microscopic theory for high-temperature superconductivity. But
making a self-consistent theoretical prediction is very meaningful and
possibly the only way to subject a particular theory experimentally testable.

A quantitative prediction of the present work is that a conserved neutral
spin current, $J_{i}^{s}$, along the $i$-th direction within the
two-dimensional (2D) plane with the spin polarized in the $\mathbf{\hat{z}}$%
-axis (out of the plane), can be generated by an applied electric field as
follows:

\begin{equation}
J_{i}^{s}=\sigma _{H}^{s}\epsilon _{ij}E_{j}  \label{current}
\end{equation}%
with $\epsilon _{ij}$ as the 2D antisymmetric tensor and $E_{j}$ the $j$-th
component of the electric field. Here the spin Hall conductivity is given by
\begin{equation}
\sigma _{H}^{s}=\frac{\hbar \chi _{s}}{g\mu _{B}}\left( \frac{B}{n_{v}\Phi
_{0}/2}\right) ^{2}  \label{sigmh}
\end{equation}%
which only depends on the intrinsic properties of the system: $\chi _{s}$ is
the uniform spin susceptibility and $n_{v}$ denotes the density of $s=1/2$
neutral spin excitations, with the electron $g$-factor $g\simeq 2,$ $\mu
_{B} $ the Bohr magneton, and $\Phi _{0}=hc/e$ the flux quantum. Note that
an external magnetic field $B$ is applied perpendicular to the 2D plane,
reducing the spin rotational symmetry of the system to the conservation of
the $S^{z}$ component only, satisfying $\frac{\partial S^{z}}{\partial t}%
+\nabla \cdot \mathbf{J}^{s}=0.$ In contrast, the charge current still
remains dissipative and the resistivity may even become divergent at low
temperature in a strong perpendicular magnetic field, leading to a spin Hall
insulator where $\sigma _{H}^{s}$ still remains finite. In this sense we may
say that the spin Hall current is dissipationless, following a similar
characterization in an anomalous Hall effect system\cite{lee}.

\begin{figure}[b]
\begin{center}
\includegraphics{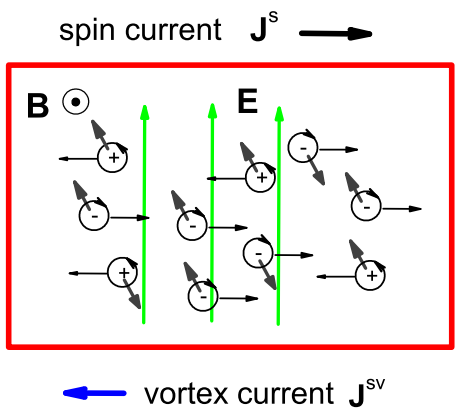}
\end{center}
\caption{(Color online) Vortices of $\pm $ vorticity can be driven by a
perpendicular electric field to form a non-dissipative current and a spin
current is simultaneously produced because each vortex is bound to an $%
S^{z}=\pm 1/2$ spin at the core in the spontaneous vortex phase (see text).}
\end{figure}

Such an SHE with conserved and \textquotedblleft
non-dissipative\textquotedblright\ spin currents is present\ in a novel
phase of the doped Mott insulator described by the phase string theory,
known as the \emph{spontaneous vortex phase }\cite{sv}, which is
characterized by a nonzero electron pairing amplitude $\Delta ^{0}$ without
true superconducting (SC)\ phase coherence: The SC pairing order parameter
is given by
\begin{equation}
\Delta ^{\mathrm{SC}}=\Delta ^{0}e^{i\Phi ^{s}}  \label{scorder}
\end{equation}%
where the phase $\Phi ^{s}$ is composed of $\pm 2\pi $ vortices which are
disordered in this state. This spontaneous vortex (SV)\ phase has been
previously proposed \cite{sv} to describe the LPP in the high-$T_{c}$
cuprate superconductors, featured by nontrivial Nernst effect \cite{ong}. In
the same theory, the Nernst signal is interpreted as contributed by the
phase slippage of vortices moving along a temperature gradient in a
perpendicular magnetic field \cite{sv}. A similar vortex state has been also
proposed \cite{anderson1} recently to described the LPP in the cuprates. But
a unique feature in the present theory is that an $s=1/2$ neutral spin
(spinon) is always trapped at the core of each vortex, as a result of the
Mott physics since the charge is depleted at the core \cite{gl,sv} [and $%
n_{v}$ in (\ref{sigmh}) is thus equal to the density of free vortices].
Because of the existence of such \textquotedblleft cheap\textquotedblright\
spinon-vortex composites (which are also present in the low-temperature SC
phase where they remain paired to ensure the SC phase coherence \cite{gl}),
an electric field can thus easily couple to the neutral spins without
involving the SO coupling to generate a spin Hall current as schematically
illustrated in Fig. 1. Therefore, the \textquotedblleft
dissipationless\textquotedblright\ spin Hall effect predicted in the present
work represents another unique property of such a phase.

\section{Spin Hall effect in a doped Mott insulator}

\subsection{Ginzburg-Landau description}

We start with a generalized Ginzburg-Landau (GL) description of the SV phase
in the phase string theory \cite{gl}
\begin{equation}
\alpha \psi _{h}+\beta |\psi _{h}|^{2}\psi _{h}+\frac{1}{2m_{h}}\left(
-i\nabla -\mathbf{A}^{s}-\mathbf{A}^{e}\right) ^{2}\psi _{h}=0~
\label{gleqn}
\end{equation}%
where $\psi _{h}(\mathbf{r})=\sqrt{\rho _{h}}e^{i\phi _{h}(\mathbf{r})}$
describes the holon condensate and the charge current is determined by
\begin{equation}
\mathbf{J}={\frac{\rho _{h}}{m_{h}}}\left[ \nabla \phi _{h}-\mathbf{A}^{s}-%
\mathbf{A}^{e}\right] ~  \label{supercurrent}
\end{equation}%
with $m_{h}$ as the effective mass. Here $\mathbf{A}^{e}$ is the
electromagnetic field, and $\mathbf{A}^{s}$ is the internal gauge field
defined by
\begin{equation}
\mathbf{A}^{s}(\mathbf{r})=\frac{1}{2}\int d^{2}\mathbf{r}^{\prime }~\frac{%
\hat{\mathbf{z}}\times (\mathbf{r}-\mathbf{r}^{\prime })}{|\mathbf{r}-%
\mathbf{r}^{\prime }|^{2}}\left[ n_{\uparrow }^{b}(\mathbf{r}^{\prime
})-n_{\downarrow }^{b}(\mathbf{r}^{\prime })\right] ~  \label{as1}
\end{equation}%
in which $n_{\sigma }^{b}(\mathbf{r})\,$\ is the spin density. Physically $%
\mathbf{A}^{s}$ depicts $\pm \pi $ fluxoids bound to spins, as
\textquotedblleft felt\textquotedblright\ by the holon condensate in (\ref%
{gleqn}), which reflects the basic mutual influence between charge and spin
degrees of freedom in the phase string theory \cite{string}.

The phase $\Phi ^{s}$ in the SC order parameter (\ref{scorder}) is related
to $\mathbf{A}^{s}$ by $\mathbf{A}^{s}=\triangledown \Phi ^{s}/2$ which
gives rise to $\Phi ^{s}(\mathbf{r})=\int d^{2}\mathbf{r}^{\prime }~\mathrm{%
Im~ln}\left[ z-z^{\prime }\right] ~\left[ n_{\uparrow }^{b}(\mathbf{r}%
^{\prime })-n_{\downarrow }^{b}(\mathbf{r}^{\prime })\right] .$ The SC phase
coherence is realized at low temperatures when all spins are
resonating-valence-bond (RVB) paired, leading to the cancellation in $\Phi
^{s}(\mathbf{r})$ \cite{gl}. Free (unpaired) $s=1/2$ spinons then give rise
to free $\pm 2\pi $ vortices in $\Delta ^{\mathrm{SC}}$ via $\Phi ^{s}$ and
thus destroy the phase coherence. It results in the SV phase with $\Delta
^{0}\propto \left( \psi _{h}^{\ast }\right) ^{2}$ still remaining finite. So
the SV phase exists in a regime $T_{c}<T\,<T_{v},$ with $T_{v}$ as the
characteristic temperature for the holon condensation$.$ Generally, at $%
T\,<T_{v}$, no \emph{free} vortices should appear in the condensate $\psi
_{h},$ except for those $\pm 2\pi $ vortices in $\phi _{h}$ whose cores are
\emph{bound} to free spinons, which then can be always absorbed into $%
\mathbf{A}^{s}$ in (\ref{supercurrent}) and $\Phi ^{s}$ in (\ref{scorder})
such that $\mathbf{A}^{s}\rightarrow \mathbf{\tilde{A}}^{s}$ and $\Phi
^{s}\rightarrow $ $\tilde{\Phi}^{s}$, with
\begin{equation}
\tilde{\Phi}^{s}(\mathbf{r})=\int d^{2}\mathbf{r}^{\prime }~\mathrm{Im~ln}%
\left[ z-z^{\prime }\right] \left[ n^{+}(\mathbf{r}^{\prime })-n^{-}(\mathbf{%
r}^{\prime })\right] ~  \label{as2}
\end{equation}%
where $n^{\pm }(\mathbf{r})=\sum_{l}\delta \left( \mathbf{r-r}_{l}^{\pm
}\right) ,$ with $\mathbf{r}_{l}^{\pm }$ denoting the coordinate of the $l$%
-th spinon carrying a $2\pi $ vorticity of sign $\pm $. So the sign of the
vorticity for a vortex carried by a spinon is not directly associated with
the spin index $\sigma $, thanks to the freedom in $2\phi _{h}$ in $\Delta
^{0}$ [or $\nabla \phi _{h}$ in $\mathbf{J}]$ which ensures the spin
rotational symmetry of the system as previously discussed in Refs. \cite%
{sv,gl}.

We now consider some important consequences of this generalized GL theory.
For a steady current state with $\partial _{t}\mathbf{J}=0,$ we find, in the
transverse gauge, the electric field $\mathbf{E}\equiv -\partial _{t}\mathbf{%
A}^{e}=\partial _{t}\mathbf{\tilde{A}}^{s}(\mathbf{r})$ in terms of (\ref%
{supercurrent})$.$ Then by using $\partial _{t}\mathbf{\tilde{A}}^{s}(%
\mathbf{r})=-\hat{\mathbf{z}}\times \pi \sum_{l}\left[ \mathbf{\dot{r}}%
_{l}^{+}\delta \left( \mathbf{r}-\mathbf{r}_{l}^{+}\right) -\mathbf{\dot{r}}%
_{l}^{-}\delta \left( \mathbf{r}-\mathbf{r}_{l}^{-}\right) \right] ,$ the
following relation can be established
\begin{equation}
\mathbf{J^{\mathrm{sv}}}=-\frac{1}{\pi }\mathbf{E}\times \mathbf{\hat{z}}%
\text{ \ \ \ }  \label{jsv}
\end{equation}%
with $\mathbf{J^{\mathrm{sv}}(r})=\sum_{l}\left[ \mathbf{\dot{r}}%
_{l}^{+}\delta \left( \mathbf{r}-\mathbf{r}_{l}^{+}\right) -\mathbf{\dot{r}}%
_{l}^{-}\delta \left( \mathbf{r}-\mathbf{r}_{l}^{-}\right) \right] $
depicting the vortex current. The Nernst signal generated by a vortex
current flowing down along the temperature gradient $-\nabla T$ has been
shown \cite{sv} based on (\ref{jsv}), which is a basic feature of the SV
phase.

In the following, we focus on the case that $\mathbf{J^{\mathrm{sv}}}$ is
driven directly by the electric field, $\mathbf{E},$ instead of by a
temperature gradient $-\nabla T,$ as schematically illustrated in Fig. 1.
Obviously, it is non-dissipative according to (\ref{jsv}) and since (\ref%
{jsv}) holds, locally individual vortices and antivortices will move in
opposite directions with $\mathbf{\dot{r}}^{+}=-\mathbf{\dot{r}}^{-}=\mathbf{%
v}^{s}$ in the uniform electric field, additively contributing to the vortex
current
\begin{equation}
\mathbf{J^{\mathrm{sv}}=}\left[ n^{+}(\mathbf{r})+n^{-}(\mathbf{r})\right]
\mathbf{v}^{s}.  \label{jsv1}
\end{equation}%
Here the densities of vortices and antivortices, $n^{\pm }(\mathbf{r}),$ is
constrained by the condition $\langle \oint_{C}d\mathbf{r}\cdot \mathbf{%
J\rangle }=0$ for an arbitrary loop $C$ such that on average
\begin{equation}
n^{+}(\mathbf{r})-n^{-}(\mathbf{r})=-\frac{B}{\pi }  \label{n+-}
\end{equation}%
based on (\ref{supercurrent}). Namely, the polarization of spinon-vortices
and antivortices is determined by the external magnetic field applied
perpendicular to the 2D plane. In the superconducting phase, without the
presence of spontaneous (thermally excited) vortices, equation (\ref{n+-})
reduces to $n^{-}=\frac{B}{\pi }$ which represents the flux quantization
condition by noting that the flux quantum $\Phi _{0}=2\pi $ in the units of $%
\hbar =c=e=1$ and the above GL theory predicts an $s=1/2$ being always
trapped at the core of a magnetic vortex \cite{gl}.

\subsection{Spin Hall effect }

Now we focus on the spins carried by these spinon-vortices. Define $%
n_{\sigma }^{\pm }(\mathbf{r})$ as the spinon-vortex density with a
vorticity $\pm $ and a spin index $\sigma $. Then $n^{\pm }(\mathbf{r}%
)=\sum_{\sigma }n_{\sigma }^{\pm }(\mathbf{r})$, and the spin current
carried by spinon-vortices can be expressed as
\[
\mathbf{J}^{s}=\frac{1}{2}\sum_{\sigma }\sigma \left( n_{\sigma
}^{+}-n_{\sigma }^{-}\right) \mathbf{v}^{s}.
\]%
Note that since the $s=1/2$ spin and the sign of the vorticity for a
spinon-vortex are independent of each other, the spin polarizations in the
magnetic field should equal for $\pm $ vortices, \emph{i.e., }$\frac{%
\sum_{\sigma }\sigma n_{\sigma }^{+}}{\sum_{\sigma }n_{\sigma }^{+}}=\frac{%
\sum_{\sigma }\sigma n_{\sigma }^{-}}{\sum_{\sigma }n_{\sigma }^{-}}.$ One
then obtains $\mathbf{J}^{s}\mathbf{=-}\frac{B}{\pi }\frac{\left\langle
S_{z}\right\rangle }{n_{v}^{2}}\mathbf{J^{\mathrm{sv}}\ }$ where $%
\left\langle S_{z}\right\rangle =\frac{1}{2}\sum_{\sigma }\sigma \left(
n_{\sigma }^{+}+n_{\sigma }^{-}\right) $ and $n_{v}\equiv n^{+}+n^{-}.$ By
using (\ref{jsv}) and $g\mu _{B}\left\langle S^{z}\right\rangle =\chi _{s}B$%
, one finally arrives at (\ref{current}) and (\ref{sigmh}) after restoring $%
\hbar $. No quantities related to dissipation explicitly appear in (\ref%
{sigmh}). Notice that both $\mathbf{J}^{s}$ and $\mathbf{E}$ are invariant
under time-reversal, and $\sigma _{H}^{s}$ is also explicitly unchanged
under $B\rightarrow -B$, in contrast to the charge Hall conductance.
Furthermore, without the spin-orbit coupling, the spin current $\mathbf{J}%
^{s}$ always remains conserved in the SV phase.

The above spin Hall effect follows directly from the GL equations [(\ref%
{gleqn})-(\ref{as1})]. Note that the non-dissipative relation (\ref{jsv}) is
independent of the \textquotedblleft Coulomb drag\textquotedblright\ between
vortices and antivortices moving in opposite directions, caused by the 2D
Coulomb-like (logarithmic) interaction which exists in the generalized GL
equations \cite{sv}. Generally, interactions do not affect (\ref{jsv}) so
that both the spinon-vortex current and spin current are \textquotedblleft
protected\textquotedblright\ in the SV phase by the pairing amplitude $%
\Delta ^{0}\neq 0$.

Then let us consider the charge current in the SV phase. Due to the presence
of free spinon vortices, the SC phase coherence is destroyed and the system
gains a finite resistivity due to the vortex motion. Such a charge
resistivity has been previously obtained based on the generalized GL theory
as follows \cite{sv}

\begin{equation}
\rho =\frac{n_{v}}{\eta _{s}}\left( \frac{\Phi _{0}}{2c}\right) ^{2}
\label{resistivity}
\end{equation}%
where $\eta _{s}$ denotes the viscosity of spinon vortices. Note that the
contribution from quasiparticles is not considered here. In the units of $%
\hbar =c=e=1,$ equation (\ref{resistivity}) can be written in a duality form
\begin{equation}
\sigma \sigma _{\mathrm{sv}}=\frac{1}{\pi ^{2}}  \label{dualc}
\end{equation}%
where $\sigma =1/\rho $ and spinon conductance $\sigma _{\mathrm{sv}}\equiv
\frac{n_{v}}{\eta _{s}}$. Thus, a charge current is always dissipative in
the SV phase, which is well separated from the spin Hall current which is
independent of the viscosity $\eta _{s}$ of spinon vortices. In this sense,
the latter is considered to be dissipationless.

\subsection{Spin Hall insulator}

In the SC phase below $T_{c}$, vortex-antivortex are bound together (spinon
confinement) and no free (unpaired) spinon-vortices present in the bulk.
Consequently, $\sigma _{\mathrm{sv}}$ vanishes such that $\sigma =\infty $
according to (\ref{dualc}). On the other hands, $\sigma $ becomes finite
when free spinons emerge in the bulk, which destroy the SC phase coherence
as discussed before and contribute to a finite $\sigma _{\mathrm{sv}}$. In
particular, if a finite density of free spinons is present at low
temperatures as stabilized by, say, a strong magnetic field, then these
unpaired bosonic spinon-vortices can experience a Bose condensation such
that $\sigma _{\mathrm{sv}}\rightarrow \infty $ at $T\rightarrow 0.$
Correspondingly, based on (\ref{dualc}), the charge conductance $\sigma
\rightarrow 0,$ leading to an insulator as the \emph{ground state} of the SV
phase. By contrast, the spin Hall conductance $\sigma _{H}^{s}$ in (\ref%
{sigmh}) remains finite as given by (\ref{uni}) below, unaffected by the
vanishing longitudinal charge conductivity. Therefore, such a ground state
of the SV phase is a \emph{spin Hall insulator,} presumably stabilized by
strong perpendicular magnetic fields.

The Bose condensation of spinons implies the existence of some sort of
antiferromagnetic ordering \cite{string}. Experimentally, both a
magnetic-field-induced magnetic ordering \cite{mag,lake} and insulating
behavior \cite{ando} have been observed in the pseudogap regime of the
underdoped cuprates. The nontrivial Nernst effect \cite{ong} and
diamagnetism \cite{ong1}, extending over a wide range of temperature, with $%
T_{v}$ as large as several times of $T_{c}$ in underdoping$,$ have been also
observed in these cuprate materials, strongly suggesting the presence of 2D
spontaneous (cheap) vortices \cite{ong,ong1,wen} as the physical origin. As
a unique prediction of the phase string theory, cheap vortices must have $%
s=1/2$ spinons located at the vortex cores due to the Mott physics which
prohibits double occupancy of electrons: a site is either occupied by a hole
or by an $s=1/2$ spin. Consequently, a dissipationless spin Hall conductance
is naturally obtained when those free vortices are either thermally excited
or magnetic-field induced in the SV phase.

Let us finally examine the magnitude of the spin Hall conductance $\sigma
_{H}^{s}$ given in (\ref{sigmh}). Generally, $n_{v}\Phi _{0}\leq B$
according to (\ref{n+-}) (the equality holds if the vortices are fully
polarized by the magnetic field) and $\sigma _{H}^{s}\leq \sigma
_{H}^{0}\equiv \hbar \chi _{s}/g\mu _{B}$. So $\sigma _{H}^{0}$ decides an
upper bound for $\sigma _{H}^{s}$. At a temperature slightly above $T_{c}$,
a typical $\chi _{s}$ can be estimated $\sim 1.1\mu _{B}^{2}/a^{2}$ $\times $
\textrm{states/eV} in the phase string theory \cite{string}, which is
comparable to the experimental values in the cuprates. By taking the lattice
constant $a\simeq 3.8\mathring{A}$, we then obtain ${\sigma }_{H}^{0}\sim
0.55(\hbar \mu _{B}/a^{2})\times \mathrm{states/eV}=0.14e$. Of course, $%
\sigma _{H}^{s}$ is generally reduced from ${\sigma }_{H}^{0}$ with the
increase of the thermally (spontaneously) excited vortices above $T_{c}$. On
the other hand, at low temperatures where the thermally excited
spinon-vortices are negligible, and all vortices are nucleated by the
applied magnetic field, one has $n_{v}=n^{-}=B/\Phi _{0}$ according to (\ref%
{n+-}). Furthermore, the spins of the vortices are totally polarized by $%
\left\langle S^{z}\right\rangle =-$ $\frac{1}{2}n_{\downarrow }^{-}=-$ $%
\frac{1}{2}n_{v}$ if the temperature is sufficiently low. Then in this limit
one finds
\begin{equation}
\sigma _{H}^{s}=\frac{1}{2\pi }e  \label{uni}
\end{equation}%
which approaches a universal number as all spins are in RVB paired except
for those associated with the vortices nucleated by the magnetic field \cite%
{gl}. Note that $\sigma _{H}^{s}=0$ if the SC phase coherence is realized
when those vortices are pinned spatially, where $\partial _{t}\mathbf{J}\neq
0$ unless the electric field $\mathbf{E}=0$ in the bulk. Namely (\ref{uni})
is valid only in the vortex flow regime at low temperature.

\subsection{Mutual Chern-Simons theory description}

So far the SHE in the SV phase has been discussed based on the generalized
GL equations (\ref{gleqn}) and (\ref{supercurrent}) of the phase string
theory. In the following we briefly discuss how this is a self-consistent
result ensured by the underlying mutual duality structure of the microscopic
theory.

The mutual-Chern-Simons effective description of the phase string theory is
given by an effective Lagrangian $\mathcal{L}_{\mathrm{eff}}=\mathcal{L}_{h}+%
\mathcal{L}_{s}+\mathcal{L}_{CS}$ \cite{qi}, with the charge part
\begin{equation}
\mathcal{L}_{h}=h^{\dagger }\left\{ i\partial _{t}-A_{0}^{s}-A_{0}^{e}-\frac{%
1}{2m_{h}}\left( -i\mathbf{\nabla }-\mathbf{A}^{s}-\mathbf{A}^{e}\right)
^{2}\right\} h  \label{lh}
\end{equation}
which describes that the charge $+e$ holon field $h$ couples to an external
electromagnetic field $A_{\mu }^{e}$ ($\mu =0,x,y$) and an internal \textrm{%
U(1)} gauge field $A_{\mu }^{s}$. The spin part $\mathcal{L}_{s}$ describes
the neutral spinon field couples to an another \textrm{U(1)} gauge field $%
A_{\mu }^{h}$, whose detailed form \cite{qi} is not important here. Here
both $A_{\mu }^{s}$ and $A_{\mu }^{h}$ can be regarded as \textquotedblleft
free\textquotedblright\ \textrm{U(1)} gauge fields, which are
\textquotedblleft entangled\textquotedblright\ by the mutual-Chern-Simons
term
\begin{equation}
\mathcal{L}_{CS}=\frac{1}{\pi }\epsilon ^{\mu \nu \lambda }A_{\mu
}^{s}\partial _{\nu }A_{\lambda }^{h}  \label{lcs}
\end{equation}%
Note that the topological constraint on $\mathbf{A}^{s}$ in (\ref{as1}) only
emerges after the temporal component $A_{0}^{h}$ is integrated out in the
partition function determined by $\mathcal{L}_{\mathrm{eff}}$. The
time-reversal, parity, and global spin rotational symmetries have been shown
to be retained in $\mathcal{L}_{\mathrm{eff}}$ at $A_{\mu }^{e}=0,$ and the
global phase diagram, including antiferromagnetic phase, SC phase, pseudogap
and SV phases, has been discussed within such a unified description \cite{qi}%
.

One can then show that $\mathcal{L}_{CS}$ will generally result in the
following equation of motion
\begin{equation}
\mathbf{J}^{s}=\frac{1}{2\pi }\mathbf{E}^{s}\mathbf{\times \hat{z}}\text{ ,
\ \ \ }\mathbf{J}=\frac{1}{\pi }\mathbf{E}^{h}\mathbf{\times \hat{z}}\text{ }
\label{eqm}
\end{equation}%
where $\mathbf{J}^{s}\equiv 1/2$ $\delta \mathcal{L}_{s}/\delta \mathbf{A}%
^{h}$ and $\mathbf{J}\equiv \delta \mathcal{L}_{h}/\delta \mathbf{A}^{s},$
with $\mathbf{E}^{s}=-\partial _{t}\mathbf{A}^{s}-\triangledown A_{0}^{s}$
and $\mathbf{E}^{h}=-\partial _{t}\mathbf{A}^{h}-\triangledown A_{0}^{h}$.
Thus, a spin current can be generated by a perpendicular \textquotedblleft
electric field\textquotedblright\ $\mathbf{E}^{s}$ and the charge current by
$\mathbf{E}^{h}$ according to (\ref{eqm}). The spin-current conservation
here is due to the \textrm{U(1) }gauge invariance associated with $A_{\mu
}^{h}$.

In particular, consider the SV phase defined by the Bose condensation of
holons with $\langle h(\mathbf{r})\rangle =\psi _{h}=\sqrt{\rho _{h}}%
e^{i\phi _{h}(\mathbf{r})}$. According to $\mathcal{L}_{h}$, $\mathbf{E}$
(via $A_{\mu }^{e}$) would accelerate the condensate unless it balanced by $%
\mathbf{E}^{s}$ (via $A_{\mu }^{s}$), satisfying $\mathbf{E}^{s}+\mathbf{E}%
=-\partial _{t}\mathbf{\triangledown \phi }_{h}+\triangledown \mathbf{%
\partial }_{t}\mathbf{\phi }_{h}$ where the right-hand-side is contributed
by the vortices in the phase of $\psi _{h}$. By incorporating the latter
into $\mathbf{J}^{s}$ which gives rise to $\mathbf{J}^{\text{\textrm{sv}}},$
similar to the previous discussion, one then reproduces (\ref{jsv}).
Therefore, the non-dissipative spin currents are quite robust so long as $%
\psi _{h}$ and thus the pairing amplitude $\Delta ^{0}$ remains finite,
which defines the SV phase. On the other hand, vacancy impurities like the
zinc substitution in the cuprates may greatly reduce the spin currents by
very effectively pinning \cite{qi1} down the free spinon-vortices. Similarly
the Nernst effect is expected to be suppressed by the same reason.

Another interesting property of (\ref{eqm}) is that a charge current flowing
through the sample will generate an \textquotedblleft electric
filed\textquotedblright\ $\mathbf{E}^{h}$, which acts on the spinon part via
$\mathcal{L}_{s}$ and thus provides a means of \textquotedblleft spin
pump\textquotedblright . It may be manipulated to design a \textquotedblleft
spin battery\textquotedblright\ in such a system \cite{sun}. Furthermore,
define $\mathbf{J}^{\text{\textrm{sv}}}=$ $\sigma _{\mathrm{sv}}\mathbf{E}%
^{h}$ and $\mathbf{J=\sigma E}$. Then by simply using the relations in (\ref%
{jsv}) and (\ref{eqm}), one easily finds the interesting duality relation (%
\ref{dualc}) between the charge conductance $\mathbf{\sigma }$ and spinon
conductance $\sigma _{\mathrm{sv}},$ previously obtained in the generalized
GL theory description \cite{gl}.

\section{Conclusion}

In the paper, the existence of a conserved dissipationless spin Hall current
is predicted in the low-temperature pseudogap regime of a doped Mott
insulator, known as the spontaneous vortex phase, based on the phase string
description. Such a spin Hall current is concomitant with the Nernst signal,
but is dissipationless whereas the latter is not. A sizable spin Hall
conductivity is obtained which only depends on the intrinsic bulk properties
as well as the external magnetic field. Furthermore, the spin Hall
conductivity remains finite even if the longitudinal charge resistivity
diverges when the pseudogap state is stabilized by strong magnetic fields in
the ground state, leading to a spin Hall insulator. Note that the spin Hall
insulator here is caused by strongly correlated effect (a quantum vortex
liquid) which has nothing directly to do with that discussed in the SO
systems where a band gap is present.

We emphasize that the spin Hall current is protected here by the
electron pairing amplitude. The SHE constitutes a sharp and very
unique prediction for the lower pseudogap phase which can be
tested in the future experiment. We point out that the present
experimental techniques \cite{exp} allow one to detect only spin
accumulations that the spin currents deposit at interfaces and
boundaries, which become detectable with the help of the strong SO
coupling \cite{exp,theo}. Thus in order to observe the SHE in the
high-$T_{c}$ materials, a natural method is to make a finite-width
strip of the relevant materials been sandwiched by semiconductors
of strong SO coupling and determine the spin accumulations at the
interfaces by the conventional techniques \cite{exp}.

Finally, if such a SHE does exist in the cuprate materials, it may have some
important potential applications in spintronics devices due to its sizable
effect, conserved and non-dissipative nature, as well as the relatively wide
region of the lower pseudogap phase. Compared to the SHE proposed \cite%
{Mur1,niu} in the semiconductor systems with SO couplings, however, we would
like also to point out a drawback in the present system, namely, a
perpendicular magnetic field is always needed here, which may hinder
practical applications and should be further carefully investigated.

\begin{acknowledgments}
We thank Z. Fang, P. A. Lee, V. N. Muthukumar, N. P. Ong, D. N. Sheng, Y.
Wang, H. H. Wen, X. G. Wen, and M. W. Wu for helpful discussions. S.P.K is
partially supported by NSFC Grant no. 10204004 and Beijing Normal
University. Z.Y.W. acknowledges supports from NSFC and MOE.
\end{acknowledgments}

\end{document}